# Reply to Comment on "Unconventional s-wave superconductivity in Fe(Se,Te)"

T. Hanaguri[1,2], S. Niitaka[1,2], K. Kuroki[2,3] and H. Takagi[1,2,4]

1.  Magnetic Materials Laboratory, RIKEN Advanced Science Institute, Wako 351-0198, Japan.
2.  TRiP, Japan Science and Technology Agency, Kawaguchi 332-0012, Japan.
3.  Department of Applied Physics and Chemistry, The University of Electro -Communications, Chofu, Tokyo 182-8585, Japan.
4.  Department of Advanced Materials, University of Tokyo, Kashiwa 277-8561, Japan.

Mazin and Singh argue that the observed peaks in the Fourier transformed spectroscopic maps in Fe(Se,Te) (*1*) may not be related to the quasi-particle interference (QPI) but would be attributed to the Bragg peaks associated with underlying chalcogen lattice and surface-induced spin-density wave (SDW) (*2*). They point out that: (i) the observed peaks at $q_2$ and $q_3$ are too sharp to be ascribed to the QPI, (ii) $q_3$ is located at the Bragg point of the chalcogen lattice, (iii) if SDW is induced at the surface and if such an SDW triggers a surface reconstruction, Bragg peak would appear at $q_2$, (iv) magnetic field would suppress both superconductivity and SDW, giving rise to the enhancement of the Bragg peak at $q_3$ at the superconducting (SC) gap energy and suppression of the Bragg peak at $q_2$, respectively. We show that these arguments are not relevant in the present case.

First, the observed peaks which have been discussed in Ref. 1 are not as sharp as Bragg peak. It is true that $q_3$ is located at the Bragg point of the chalcogen lattice but the QPI signal is distinct from the lattice Bragg peak. In Fig. 1, we show linecuts from the Fourier-transformed conductance-ratio map $Z(\mathbf{q}, E)$, in which QPI peaks appear (*1*), along the line which passes both $q_2$ and $q_3$. Linecut from the Fourier-transformed topographic image (Fig. 1A of Ref. 1) is also shown to give an idea of the sharpness of the Bragg peak. In the absence of magnetic field (black lines), both peaks at $q_2$ and $q_3$ in $Z(\mathbf{q}, E)$ are much broader than the Bragg peak. Bragg-like sharp feature emerges at $q_3$ at high energies but near the SC-gap energy (1 ~ 3 meV), only broad feature dominates. Indeed, the widths of the peaks are comparable to 20 % of the Brillouin zone dimension of $2\pi/a$ where $a$ is the inter-chalcogen distance (an arrow in Fig. 1A), as suggested by Mazin and Singh (*2*).

When magnetic field is applied (red lines in Fig. 1A), Bragg-like sharp feature grows at $q_3$. Note that the field enhancement of this Bragg-like peak persists well above the SC-gap energy, which clearly suggests that the enhancement can not be explained by the suppression of the SC quasi-particle peak alone. On the contrary, pre-existing broad peak is strongly enhanced only near the SC-gap energy, suggesting that it is related to superconductivity. Namely, features at $q_3$ consist of two components, a sharp Bragg-like peak and a broad peak. What we ascribed to the QPI peak in Ref. 1 is the latter.

Surface reconstruction triggered by surface-induced SDW is an interesting proposal. However, there is no evidence that such a reconstruction or SDW are really induced at the surface of Fe(Se,Te). Because the cleaved surface of Fe(Se,Te) is neutral, it may be electronically more robust than the surfaces of other iron-based superconductors, e.g. $AFe_2As_2$ (A: alkali-earth element), which are inevitably charged. Even if SDW would be induced at the surface, relation between the SDW and the peak at $q_2$ is a highly non-trivial issue. Because charge amplitude induced by SDW does not depend on the spin direction, neither $\sqrt{2}\times\sqrt{2}$ nor $2\times1$ SDW would generate charge superstructure. It is not straightforward that such a charge-uniform state triggers a surface reconstruction with a charge superstructure. In any case, the observed peaks are broad enough to discard the Bragg-peak scenario and are consistent with the QPI peak.

QPI experiment is an emergent technique and full theoretical analysis is still underway even in a case of cuprate (*3-6*). We anticipate that further experimental and theoretical works will shed more light on QPI in more complicated compound like an iron-based superconductor.

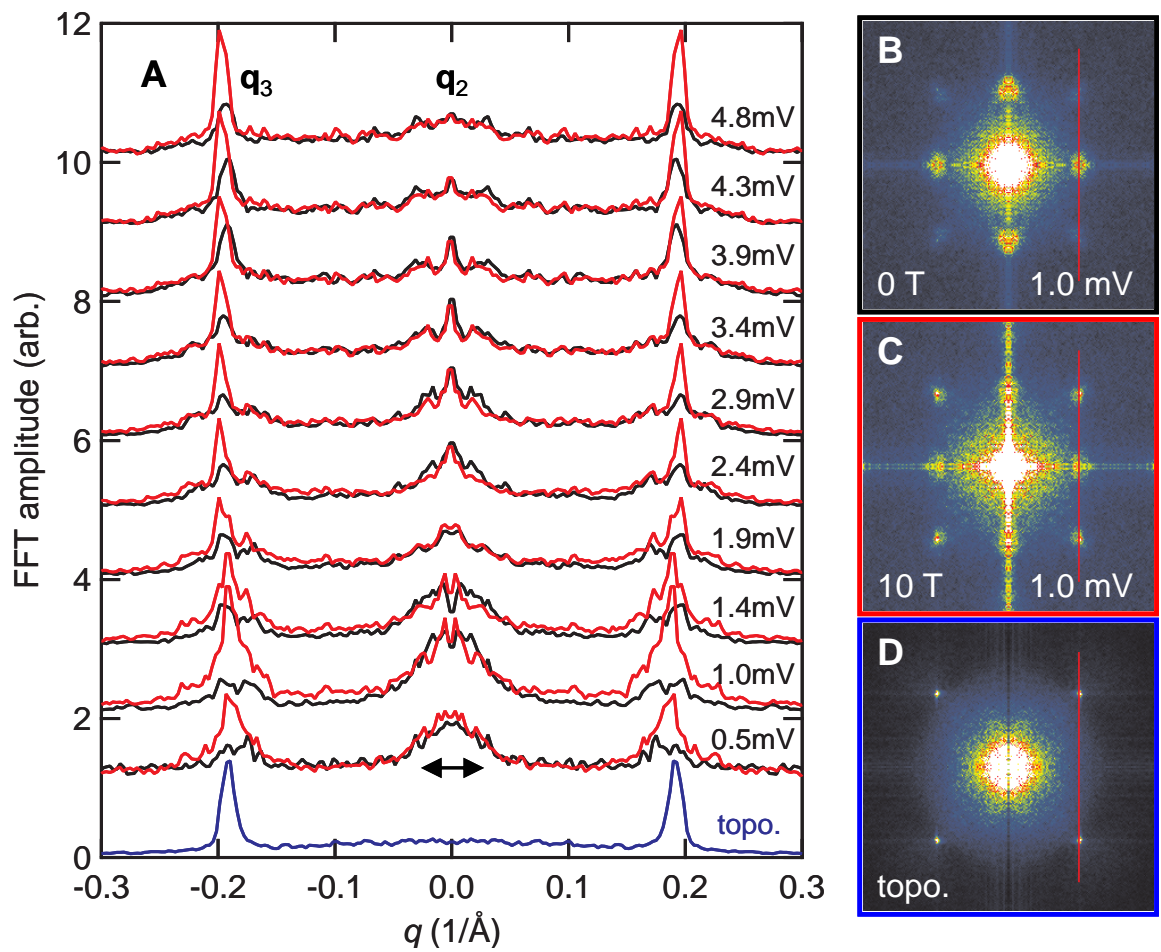

**Fig. 1.** (**A**) Linecuts from $Z(\mathbf{q}, E)$ maps and Fourier-transformed topographic image along the line shown in (**B**)-(**D**). Black and red curves denote the data taken at 0 T and 10 T, respectively. An arrow indicates 20 % of $2\pi/a$.